\documentstyle[12pt,aaspp4]{article}

\begin{document}

\centerline{ To be published in:}
\centerline{\bf ``Astrophysics with Infrared Arrays: A Prelude to SIRTF"}
\centerline{\it ASP Conference Proceedings }

\title{Review:  Understanding Galaxy Formation and Evolution with Long Wavelength Observations}

\author{Matthew A. Malkan}
\affil{U.C.L.A. Astronomy, Los Angeles CA 90095-1562}

\begin{abstract}
It is well established in the local Universe that regions of high
star-formation rate are dusty.  As a result of this physical
causal link, galaxies of increasing current star formation activity
emit a larger proportion of their bolometric luminosity via
dust absorption and re-radiation in the thermal mid- to far-infrared.
Several new observations indicate that this trend continues back to earlier
cosmic times, when star formation rates were very high, and a large
fraction of the resulting UV power was reprocessed by dust into the
infrared.  For studying the more luminous, (and therefore more dusty)
galaxies, infrared spectroscopy is crucial even at moderate redshifts.
For the less dusty galaxies, we are still driven strongly to longer wavelengths
as their redshifts increase from 5 to 10.  On the observational side, infrared astronomy is about
to start catching up with optical astronomy in sensitivity, spatial resolution and field-of-view.  Therefore, long-wavelength observations will play
the key role in our understanding of galaxy formation and evolution
in the ``Immature Universe."
\end{abstract}

\section{Dust in the Universe: Now}

Figuring out how galaxies formed and evolved a long time ago is
one of the supreme challenges for theorists and, especially, observers.
We will need all the help we can get.
As a great coach might have said: ``When the going gets tough, the
tough get empirical."  It is good advice to review our empirical
knowledge of star-forming
galaxies in the local Universe if we want to have a good shot at
understanding them in the early Universe.
Since the highly successful IRAS mission, 
no one (at least in astronomy) can be ignorant of the fact
that many galaxies are moderately to extremely dusty. 
Operationally, this means that a significant, to a dominant fraction of
their bolometric energy is reradiated at long wavelengths after absorption
by dust grains.

Many astronomers are also aware of the next lesson from IRAS: 
the effects of dust are strongest in the most luminous galaxies.
As the current star formation rate increases, so does the fraction of stellar
photons that are absorbed by dust.  This could have been predicted from the
simple fact that young stars are generally closer to the dusty molecular
clouds out of which they were born than are older stars.
So naturally the youngest stars have the highest ``dust covering fractions",
and this trend is strong when integrated over a galactic scale.

This is shown in Figure 1, reproduced from Malkan and Stecker 1998.
It simply presents the average observed far-infrared-to-optical spectra of
galaxies of varying luminosities, based on the large multiwavelength
database of Spinoglio et al. (1995). The lines are average spectra of galaxies
differing by factors of ten in bolometric luminosity.
At higher luminosities, the ``valley" around log $\nu=13$, 
between the light of red giant
stars on the right (peaking around log $\nu=14.5$) and the cold dust
bump on the left (around log $\nu=12.5$) is progressively ``filled in"
from more and more thermal emission from warm dust grains 
associated with H II regions.  And the relative power in the dust
emission grows to be much stronger than that of the starlight,
as the dust peak shifts to shorter wavelengths.
At the same time, the relative power at short wavelengths
(log $\nu=15$, off the right end of the graph) 
is collapsing due to dust absorption.
The dominant factor in making galaxies in the local universe more
luminous is not their larger masses, but their lower Mass/Light ratios--
the result of a higher proportion of recently formed stars.

In the extreme ultraluminous starburst galaxies, the average dust absorption
becomes so large that much of the galaxy is effectively optically thick
at wavelengths of 10$\mu$m or longer.
ISO is providing dramatic new evidence for this proposition.
Figure~2
shows a montage of Long Wavelength Spectrometer grating-scan
spectra of luminous galaxies from the LWS Infrared-Bright Galaxies
team (Fischer et al. 1998).  Even at the resolution 
of about 200, the strong forbidden lines from neutral and ionized gas
are evident in the normal galaxies.  Most of this emission becomes
weaker {\it relative to the thermal dust continuum} in the more
luminous objects, while at the same time molecular {\it absorption}
lines become more and more prominent. For the molecular lines, and even
[OI] 63$\mu$m to be detected in absorption requires very high
column densities of cold gas.
There is a tendency for dust absorption to be stronger in the more
luminous galaxies, but it is not strictly  monotonic.

\section{Dust in the Universe:  Then}

It has been said that the Universe is difficult to understand because
``there's nothing to compare it with."
Much accumulating observational evidence indicates that we {\it can}
compare the Immature Universe with our current epoch and locale--
the same trends we find at z=0 continue
at high-redshifts. The main observational evidence for this view is:

\begin{itemize}
\item a) {\bf 
Detections of distant dusty galaxies 
in 
ultra-deep ISO fields;}
It is always scary when unexpected sources are detected right
at the sensitivity threshold of new instruments, and this concern hovers
around most of the evidence to be described here.  Nonetheless, 
several groups have now begun to converge on estimates of very deep
infrared galaxy counts, using ISOCAM at 7 and 15$\mu$m (Oliver, et al. 1997;
Taniguchi, et al. 1997; Desert, et al. 1998),
and ISOPHOT at 175$\mu$m (Kawara, et al. 1998).
The surface densities are relatively high: 0.7 to 1.7 per square arcmin
in the ISOCAM images and 0.011 per square arcmin in the ISOPHOT map.
Especially at the longer wavelengths, many of the  
faint sources are probably  
starburst galaxies at redshifts of 1 or greater.  High rates of
star formation are implied by the required high luminosities in
the mid-infrared.
Whatever the exact redshifts of these distant galaxies are, 
in many cases their {\it ratio}
of infrared/optical luminosity is so large, that it requires
over half of their power to be reprocessed by dust grains (Rowan-Robinson et al.
1997).  The large number of detections (for such small areas covered 
so far) suggests that this may be
the rule rather than the exception for high-redshift galaxies.
The longer wavelength observations also highlight the importance of
observing regions of low cirrus contamination from the Milky Way.
Fluctuations in the column density of local interstellar matter on
sub-arcminute scales are a significant, if not dominant source of noise
at these faint levels.
The real solution, however, lies with improved spatial resolution from 
larger telescopes, which is also essential for confident cross-identifications
at other wavelengths.

\item b) {\bf High surface densities of star-forming galaxies in narrow-band
infrared searches;}
Figure~3 shows the estimated star-formation rates 
from recent detections of $H\alpha$ emitting
galaxies at redshifts of 2 to 2.5 from Teplitz, Malkan and McLean (1998a).
The same narrow-band imaging technique has also recently been used
to detect a [OIII]$\lambda$5007 line-emitting galaxy at z=3.31 (Teplitz,
Malkan and McLean 1998b).  

Table 1 lists the implied surface density of galaxies 
per square arcminute, for comparison
with other search methods, but it is also necessary to account for
the very different redshift windows of each one.
Although the surveys are not yet large, and the fluctuations from
one to another are substantial, it appears that near-infrared
selection finds several times more galaxies than optical methods,
which are based on detecting spectral features in the rest-frame ultraviolet.
The highly successful Lyman break search technique can only find galaxies that
have 1) {\it blue continuum colors.} Most of the confirmed U-band dropouts
in the HDF have V-I$\le$0.5 (e.g. Dickinson, 1998); and 
2) {\it strong far-UV continuum.}  If Ly$\alpha$ is not strong in emission, (and
this can be prevented by very minor amounts of dust), then the spectroscopic
confirmations hinge almost completely on identifying  key 
interstellar {\it absorption} 
lines between 1200 and
1400$\AA$.

 Is it reasonable to suppose that a major proportion 
(half or even more?)
of high-redshift galaxies are missed by optical searches, but detectable
in the near-infrared?  Yes, especially when one considers that the
integrated luminosity  of
those galaxies {\it detected} in the optical surveys
(i.e. the {\it least} dusty ones) 
needs to be corrected upward for extinction by 
a factor of 
five to seven 
(Dickinson 1998; Pettini et al. 1998).    
Some studies have suggested even larger corrections (Meurer, et al. 1997;
Sawicki and Yee, 1998).  
On the other hand, extensive spectroscopy of a K-magnitude-limited
galaxy sample has been used to argue that the correction should
be only a factor of three, and that the incompleteness is not extremely
large (Cohen et al 1998). 
There are tentative indications that the effects of extinction
are stronger in the more luminous galaxies.   This question will be answered definitively when near-infrared
spectrographs on large telescopes  (NIRSPEC on Keck, for example)  can measure the rest-frame optical spectra of
high-redshift galaxies as accurately as current optical spectrographs are measuring
their rest-frame ultraviolet spectra.
\begin{table}
\caption{Surface Densities of High-Redshift Galaxies.} \label{tbl-1}
\begin{center}\scriptsize
\begin{tabular}{crrrrrrrrrrr}
Surf.Density\tablenotemark{1} & Limiting Flux\tablenotemark{2}& Survey & Targeted Where? & Avg. z & $\Delta$z & \\ 
\tableline
1.2 &0.5--0.8 &TMM 98a &QSO ABSN  & 2.3--2.5  & 0.04 \\
0.5 &2.0 & Bechtold et 98  &DLA  & 2.3 & 0.04 &\\
0.035 & 2.0 & Bunker et 95 & DLA & 2.3 & 0.04 & \\
0.004 & 3.0+ & Beckwith et 98 & QSO Ze &2.3 & 0.04 & \\
0.15  & 3.0- & Mannucci et 98  & DLA   & 2.3 & 0.04 & \\
0.3 & 0.3--0.4  & TMM 98b & QSO Ze & 3.3  & 0.04 & \\
\tableline
1   & 1?  &  UV Dropouts & Field & 2.8--3.3 & 0.5 \\
0.00001 &  & RMS 93  & IRAS All Sky & 0--0.04  &  0.02 \\
0.0001 &  &  Gallego et al & Field & 0--0.1 & 0.04 \\
\tableline
\end{tabular}
\end{center}
\tablenotetext{1}{Objects per square arcminute}
\tablenotetext{2}{in units of $10^{-16} erg/sec/cm^2$}
\end{table}
\item c) {\bf Detection of strong Diffuse Infrared Background (DIRB) emission}
(that requires substantial luminosity evolution in the
infrared);
Several recent calculations have attempted to predict the DIRB
emission from galaxies.  Figure~4  shows one of the relatively 
conservative estimates, based on a simple {\it empirical} assumption:
that the trend for more luminous galaxies to emit a larger fraction
of their power in the thermal infrared continues back to z=1--2,
when galaxies were more than an order of magnitude more luminous than
today. 
The smooth curves are the predictions from Malkan and Stecker (1998),
which extrapolate from the best present-day IRAF luminosity functions.
Note that the predictions are very similar using either the IRAS
luminosity functions at 12$\mu$m (upper dashed line, excludes Seyfert
galaxies) or at 60$\mu$m
(upper solid line, includes Seyfert galaxies). 
The two upper solid lines assume the best guess
evolution law inferred from deep IRAS
counts, $L \propto (1+z)^3$, out to z=2 (upper curve) or z=1 (next upper solid
line).  The lower lines are conservative lower limits 
assuming small luminosity evolution.
The error bars show new claimed {\it detections}, 
from the DIRBE and FIRAS experiments
on COBE and from ISOCAM number counts described above 
(Fixsen et al. 1998; Puget et al. 1996; Dwek et al. 1998).
We see that our empirical assumption
is broadly consistent with the reported detections at 7--500$\mu$m.
If the new DIRBE 
detections at 140 and 240$\mu$m are confirmed (they are formally
consistent with the lower FIRAS estimates (triangular region),
given the large error bars), 
they will require
even {\it more} dusty galaxies than are included in our simple
empirical model. The wavelength of this exess is sufficiently longward of
the peak in low-redshift galaxies that for nearly any model it requires a large
population of dusty galaxies at redshifts of 1 or more.
(Traditional active galactic nuclei--Seyfert 1's and 2's-- 
make a small (10\%)
addition to these totals--shown by the dotted lines-- 
but they are not sufficiently numerous to
alter the overall conclusions significantly.)
Independent calculations (e.g., Guiderdoni et al. 1998) lead to
similar conclusions.

\item d)  {\bf Millimeter line and continuum detections of distant galaxies.}
In a few systems where a distant background object
(e.g., a Seyfert galaxy: FSC10214+4724, or a quasar: H1413+117, the Cloverleaf)
is gravitationally lensed, sensitive searches have detected
molecular emission lines
(Barvainis et al. 1997; Downes et al. 1995: Frayer et al. 1998).
High-redshift quasars, or galaxies grouped with them, have also
been detected in the 1.3mm continuum and CO line emission
(Ohta et al. 1996; Omont et al. 1996a; Omont et al. 1996b).  
It is generally assumed that
the millimeter line and continuum emission is associated with star formation
in the host galaxy, and is not related to the active nucleus.
If so, and if these objects are
representative of many high-redshift galaxies, then dust grains
and the molecules that they help to form may be common at high redshifts.

\item e) {\bf Detection of a high surface density of faint submillimeter
sources by SCUBA on the JCMT.}  Making deep source counts (to mJy levels at
850$\mu$m) was one of the chief goals for building SCUBA. 
Some of its first day-long pointings under good sky conditions were made by
Smail, et al. (1997), who targeted 
Abell clusters, in the hope that they would gravitationally
amplify the faint background sources they were seeking.
They were rewarded with detections which imply 0.6 sources per
square arcminute down to 4 mJy, though it is not certain
how strong the lensing effects were.
Hughes et al. (1998) have 5 detections in the Hubble Deep Field,
and Barger et al. (1998) have one detection each in the Lockman
Hole and in SSA13.  
As with ISO, there is some judgement
in deciding what to identify as a significant detection, but most
groups appear to be proceeding fairly cautiously. 
It is the same situation as in the other surveys described above:
one might not have too much confidence in the results of just one
or two deep fields, but as different groups, independently studying different
regions, continue to find comparably large densities of faint
sources at 850$\mu$m, the significance of the conclusion builds.
\footnote{I am assuming there are {\it not} many equally sensitive
(unpublished) observations with {\it no} detections (i.e. that we are
not just sleeping through a Bayesian nightmare).}
Assuming some of these sources are at cosmologically interesting
distances, and that most of their continuum emission is powered
by star formation rather than a nonstellar AGN, they must 
be the tip of the young galaxy iceberg: tremendous starbursts with SFR's 
of hundreds to a thousand $M_{\odot}$ /year (e.g., Franceschini et al. 1998).
Major improvement in the number statistics will require larger detector
arrays; pushing to deeper sensitivities will 
probably require interferrometry, to avoid the
looming confusion limit (Blain, et al., 1998).
In any case, some near-IR spectroscopy will again be required,
to confirm their redshifts, and to prove that they are not powered by
AGN.
\end{itemize}

\section{``In Today's Rapidly Evolving World of the Future:"
Bright Prospects for Coming Infrared Observatories in Space}

Based on the above considerations, I can confidently make this
prediction:
In the upcoming decade, infrared/long-wavelength astronomy(s) 
\footnote{ In fact, the long-wavelength observational techniques will
soon become so necessary in this field that the phrase ``infrared astronomer"
will soon be an anachronism.  Instead, there will be ``observers" who
study high-redshift galaxies, and it will simply be assumed that 
infrared instruments are a prime part of their arsenal.}
will play the leading
role in advancing our understanding of galaxy formation and evolution.
The support for this conclusion gets stronger at every successive
(almost quarterly!) conference on anything containing the phrase
``Young Universe", including the October 1997 Monteporzio 
meeting with that name.  The prediction stands on two legs:
1) where the most observational gains are; and
2) where the distant sources emit.
The first point, the {\it unique} instrumentation gains that infrared
astronomers are on the verge of reaping 
(as they in effect ``catch up" with more mature
wavebands such as the optical), are widely appreciated, but are
so dramatic they merit re-summarizing:  

\begin{itemize}
\item 1) orders of magnitude gains in sensitivity, particularly from
airborne and space-based telescopes with greatly reduced sky backgrounds;
\item 2) orders of magnitude increases in the number of pixels,
which will be particularly useful to exploit the next gain:
\item 3) orders of magnitude improvement in spatial resolution,
often pushing right up to the diffraction limit of the telescope
(or beyond in the case of sub-millimeter interferrometers); combined with
\item 4) favorable K corrections, at least relative to shorter wavelengths.
At the longer wavelengths and high
redshifts, the K corrections actually can become zero, or even (longward
of 60$\mu$m) {\it positive.}
\end{itemize}

Even beyond all these fabulous gains, 
it is a truism that we need to study red objects at red wavelengths.
Many of these distant galaxies are largely or totally opaque at rest
wavelengths shortward of several microns, and nearly all of
their power emerges at 2--100$\mu$m.  Given their huge visual extinctions,
their most useful diagnostic
emission lines from their neutral and ionized gas have rest
wavelengths of 2$\mu$m to longward of 35$\mu$m.
By far the strongest spectroscopic features (in equivalent width)
are the PAH bands at 3.3, 7.7 and 11.3$\mu$m (not included in the calculations
of Figures 1 and 4).
The strongest emission line at long wavelengths, [CII]158$\mu$m
(e.g., Colbert et al. 1998), will be detectable at {\it high} redshifts
by upcoming ground-based millimeter arrays.

To take a specific example, for each galaxy we can use the
ratios of infrared emission lines to estimate what proportion
of the bolometric energy output comes from young stars, and what
proportion comes from a (possibly hidden) active galactic nucleus.
Both processes coexist in low-redshift galaxies (such as NGC 1068 and
NGC 7469), and also in high-redshift galaxies (e.g. Malkan, Teplitz
and McLean 1996; Frayer et al. 1998).  
The strength of high-ionization forbidden fine-structure
emission lines from 2 to 50$\mu$m is predicted to be a powerful
indicator of the relative contributions of starbursts and AGNs
(Spinoglio and Malkan 1992).
Recent ISO spectroscopy with the Short Wavelength Spectrometer shows
that this can be done in practice (see review by Genzel in this
proceeding).

To what redshift will we continue to find many dusty
galaxies emitting strongly in the infrared? The answer depends on the 
trade-off between two opposing effects:

\begin{itemize}
\item 1) the higher past star formation rates, which are empirically
linked with a larger bolometric fraction of power reprocessed
by dust grains (discussed above), {\it VERSUS}
\item 2) the tendency for more metal-deficient systems to be less dusty.
We do not know how strong this trend is, even in the local Universe.
We do know that most visible parts of the Immature Universe became
enriched with at least some metals (up to Population II abundances
of 1 to 10\% solar) in a short time.  This is merely a restatement of
the ``G dwarf" problem in the Milky Way (the virtual absence of
extremely metal poor old stars), or the absence of direct evidence
for the fabled ``Population III."  The evolutionary phase before any dust
grains had a chance to form could be extremely brief.
\end{itemize}
My 
suspicion is that 1) will tend to dominate 2) at least
back to a redshift of 5 to 10.

Even at the highest redshifts when the first galaxies were forming,
infrared spectroscopy will be absolutely essential to study the
fraction of young galaxies which will be less dusty than the
luminous IR galaxies.  Their  strongest, most powerful spectroscopic
features, in emission and absorption, start at [OII]3727
and extend up to the CO absorption and H recombination lines
in the 2$\mu$m window.  More and more of these key features are shifted
longward of $\lambda_{obsvd} \sim 5\mu$m at cosmologically interesting lookback
times (z increasing from ~2 to $\ge$ 10) and most of them will not
in general fall 
into any of the atmospheric "windows" for ground-based telescopes.

As usual, spectroscopy
\footnote{ A spectral resolution of a few hundred km/sec is
reasonably matched to the intrinsic width of
most of these features.  Even when they are unresolved,
e.g. in slitless spectroscopy grism surveys, they are
still extremely useful for answering most of the above
questions.}
of the strongest features
is the single most important tool available
for studying these galaxies.  It is needed to 
identify them and measure their redshifts,
measure the kinematics, masses 
and dynamics of their stars and interstellar medium,
determine the ionization and excitation mechanism(s) in their gas
(i.e., active nucleus, young stars, shocks, etc.)
This then allows estimates of elemental abundances,
and ultimately determination of their evolutionary status.

For these reasons, much of the action in studies of the
``Immature Universe" for
the next 10 years will come from long-wavelength spectroscopy,
especially with SOFIA and SIRTF.  Those same reasons  
make an overwhelmingly strong case
for extending the capabilities of a Next Generation Space Telescope
beyond 5$\mu$m, preferably to 35$\mu$m.

P.G. Wodehouse's immortal Bertie Wooster summed up the current situation
best, when he exclaimed:
``Half of the world has {\it no idea} how the other two-thirds lives!"
This is going to change dramatically in the next few years.


\bigskip
\bigskip

I wish to thank the colleagues who shared their work on ISO data in
advance of publication.  ISO research at UCLA has been supported by
NASA grant NAG 5-3309.
Harry Teplitz also provided help with Figure 3.

\vskip 1.0 in



\begin{figure}
\plotone{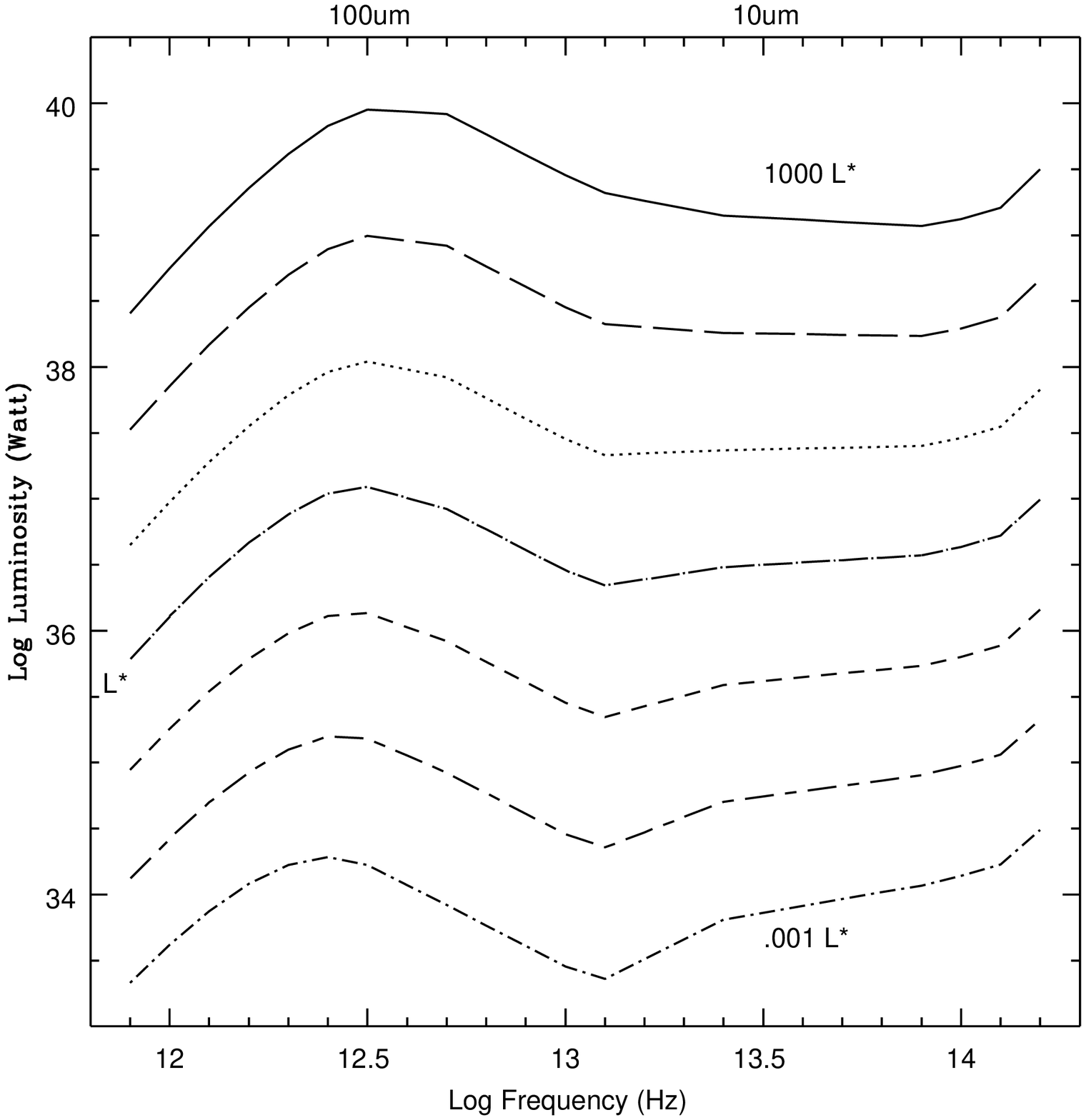}
\caption{Average Observed Infrared Spectra of Galaxies of Varying Luminosity (Malkan and Stecker 1998).}
\end{figure}

\begin{figure}
\plotone{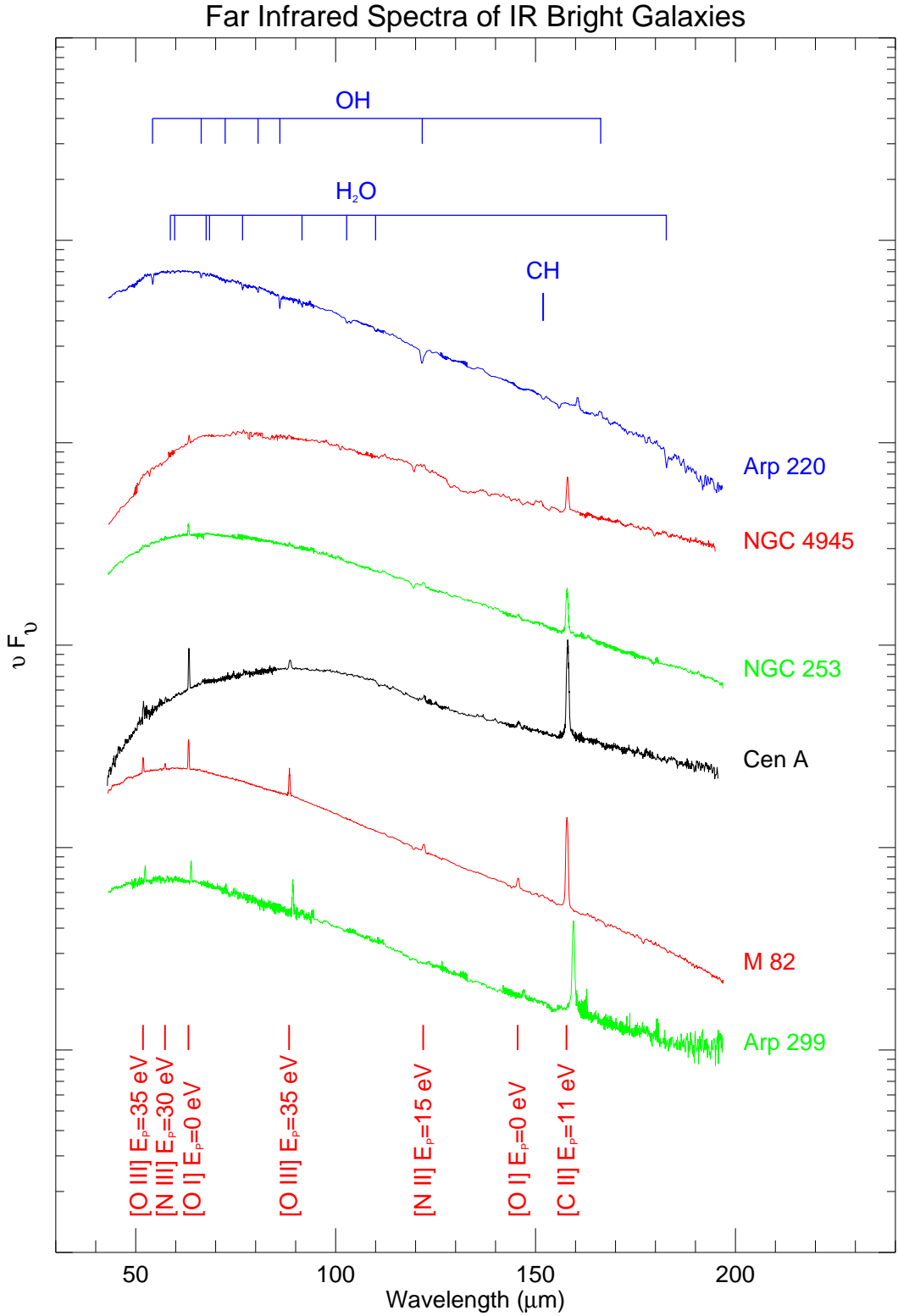}
\caption{ISO LWS Full Grating Scan Spectra of IR-Bright Galaxies (Fischer et al. 1998).}
\end{figure}

\begin{figure}
\plotone{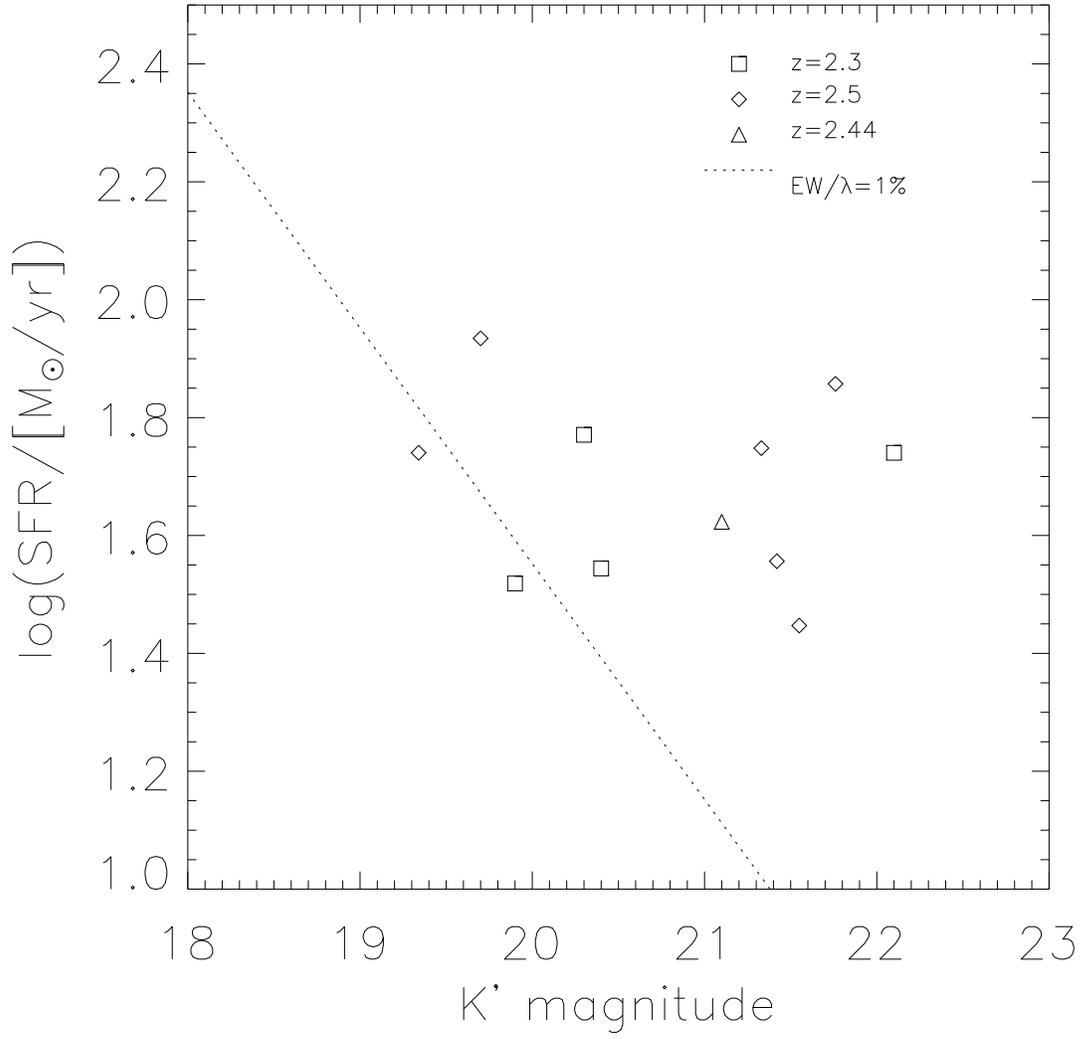}
\caption{H$\alpha$-emitting galaxies detected at z=2.3--2.5 (TMM 98a).} 
\end{figure}

\begin{figure}
\plotone{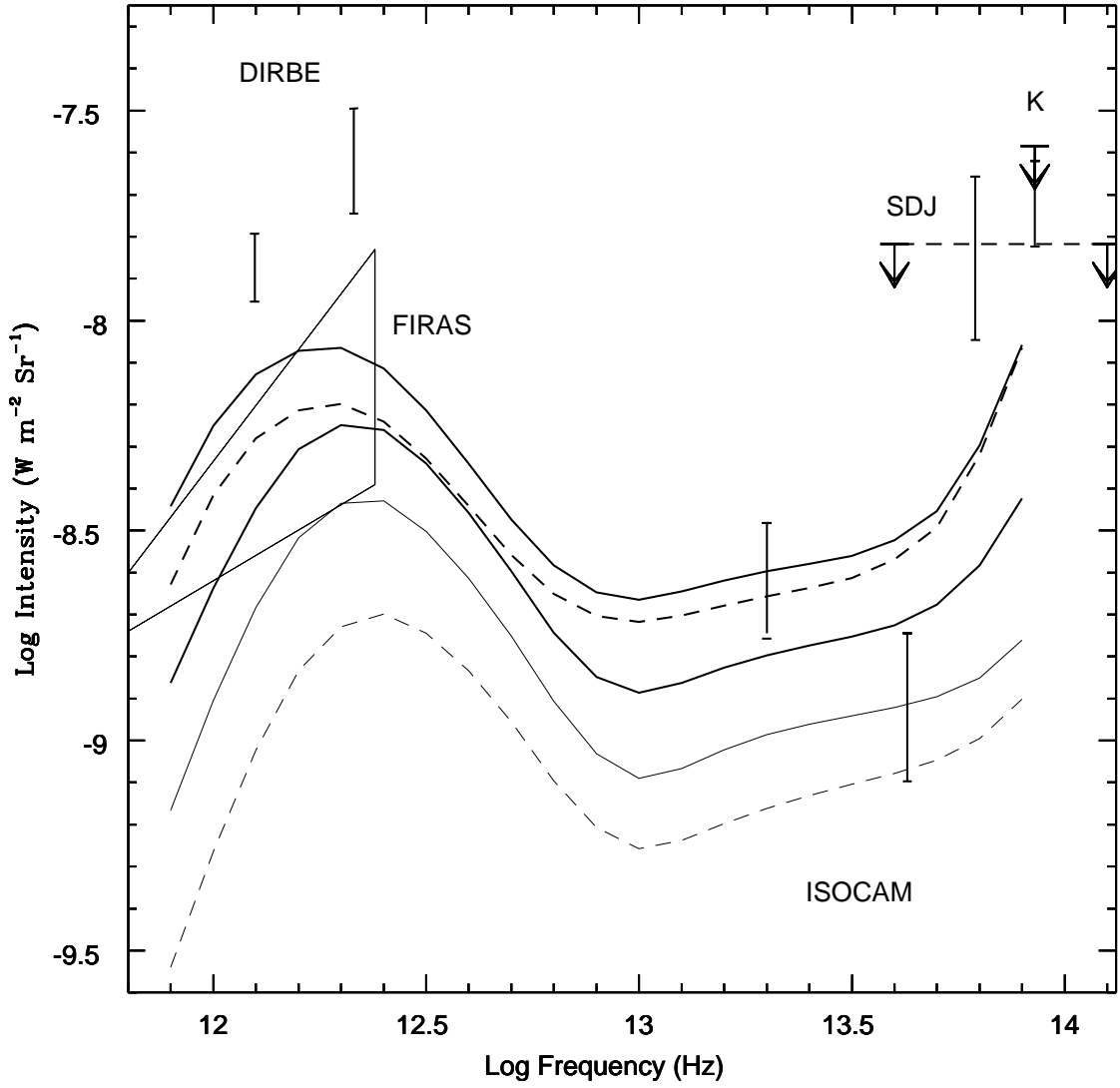}
\caption{Estimates of the Cosmic Diffuse Infrared Background Radiation: Models and Data.} 
\end{figure}
\end{document}